# MacWilliams Type identities for $m$-spotty Rosenbloom-Tsfasman weight enumerators over finite commutative Frobenius rings[*]

Minjia Shi

*School of Mathematical Sciences of Anhui University, 230601 Hefei, Anhui, China*

**Abstract** The $m$-spotty byte error control codes provide a good source for detecting and correcting errors in semiconductor memory systems using high density RAM chips with wide I/O data (e.g. 8, 16, or 32 bits). $m$-spotty byte error control codes are very suitable for burst correction. M. Özen and V. Siap [7] proved a MacWilliams identity for the $m$-spotty Rosenbloom-Tsfasman (shortly RT) weight enumerators of binary codes. The main purpose of this paper is to present the MacWilliams type identities for $m$-spotty RT weight enumerators of linear codes over finite commutative Frobenius rings.

## 1 Introduction

The error control codes have a significant role in improving reliability in communications and computer memory system [2]. For the past few years, there has been an increased usage of high-density RAM chips with wide I/O data, called a byte, in computer memory systems. These chips are highly vulnerable to multiple random bit errors when exposed to strong electromagnetic waves, radio-active particles or high-energy cosmic rays. To make these memory systems more reliable, spotty [17] and $m$-spotty [13] byte errors are introduced, which can be effectively detected or corrected using byte error-control codes. To determine the error-detecting and error-correcting capabilities of a code, some special types of polynomials, called weight enumerators, are studied.

In general, the weight enumerator of a code is a polynomial describing certain properties of the code, and an identity which relates the weight enumerator of a code with that of its dual code is called the MacWilliams identity. Recently various weight enumerators with respect to $m$-spotty Hamming Weight (Lee weight and RT weight) have been introduced and studied for various types of codes. With respect to $m$-spotty Hamming Weight, Suzuki et al. [12] defined Hamming weight enumerator for binary byte error-control codes, and proved a MacWilliams identity for it. M. Özen and V. Siap [6] and I. Siap [9] extended this result to arbitrary finite fields and to the ring $\mathbb{F}_2 + u\mathbb{F}_2$ with $u^2 = 0$, respectively. In a previous work, we generalized the result in [9] to ring $\mathbb{F}_2 + u\mathbb{F}_2 + \cdots + u^{m-1}\mathbb{F}_2$ with $u^m = 0$ in [11]. I. Siap [10] defined $m$-spotty Lee weight and $m$-spotty Lee weight enumerator of byte error-control codes over $\mathbb{Z}_4$ and derived a MacWilliams identity for the same. A. Sharma and A. K. Sharma introduced joint $m$-spotty weight enumerators of two byte error-control codes over the ring of integers modulo $\ell$ and over arbitrary finite fields with respect to $m$-spotty Hamming weight [16], $m$-spotty Lee weight [14] and $r$-fold joint $m$-spotty weight [15]. They also discussed some of their applications and derived MacWilliams type identities for these enumerators. M. Özen and V. Siap [7] proved a MacWilliams identity for the $m$-spotty RT weight enumerators of binary codes. In this paper, we will consider MacWilliams type identities for $m$-spotty RT weight enumerators of linear codes over finite commutative Frobenius ring, which generalizes the result of [7]. The organization of this paper is as follows: Section 2 provides definitions of $m$-spotty RT weight and $m$-spotty RT distance. Section 3 presents MacWilliams type identities for $m$-spotty RT weight, and Section 4 illustrates the weight distribution of the $m$-spotty byte error control code by two examples. Finally, the paper concludes in Section 5.

## 2 Preliminaries

---

[*]E-mail address: smjwcl.good@163.com (Min-Jia Shi).
**The original manuscript was submitted for reviewing on 2nd November 2012.**
This research is supported by NNSF of China (61202068, 11126174), Talents youth Fund of Anhui Province Universities (2012SQRL020ZD).

In this section, we begin by giving some basic definitions that we need to derive our results. Let $\mathbb{R}_\ell$ be a finite commutative Frobenius ring with unity and $N$ be a positive integer. Let us recall some basic knowledge about $\mathbb{R}_\ell$ as describe in [3]. Writing the identity element 1 of the ring as the sum of the primitive idempotents of $\mathbb{R}_\ell$, we obtain an isomorphism

$$\mathbb{R}_\ell \xrightarrow[\phi]{\cong} R_1 \oplus \cdots R_s,$$

where $R_1, \cdots, R_s$ are local commutative rings. The finite commutative ring $\mathbb{R}_\ell$ is called a *Frobenius ring* if $\mathbb{R}_\ell$ is self-injective (i.e., the regular module is injective), or equivalently, $(C^\perp)^\perp = C$ for any submodule $C$ of any free $\mathbb{R}_\ell$-module $\mathbb{R}_\ell^n$, where $C^\perp$ denotes the orthogonal submodule of $C$ with respect to the usual Euclidean inner product on $\mathbb{R}_\ell^n$. Moreover, in this case, $|C^\perp||C| = |\mathbb{R}_\ell|^n$ for any submodule $C$ of $\mathbb{R}_\ell^n$, where $|C|$ denotes the cardinality of $C$. This is one of the reasons why only finite Frobenius rings are suitable for coding alphabets. With the isomorphism $\phi$, $\mathbb{R}_\ell$ is Frobenius if and only if every local component $R_i$ is Frobenius, and the finite local Frobenius ring $R_i$ is Frobenius if and only if $R_i$ has a unique minimal ideal. Note that, in the non-commutative case, a self-injective ring is called a *quasi-Frobenius ring*, while one more condition is required for it to become a Frobenius ring. However, in the commutative case, a finite quasi-Frobenius ring is exactly a finite Frobenius ring. The reader may refer to [18,19] for more details on Frobenius rings.

Let $\mathbb{R}_\ell^N$ be the $\mathbb{R}$-module of all $N$-tuples over $\mathbb{R}_\ell$. For a positive divisor $b$ of $N$, a byte error-control code of length $N$ and byte length $b$ over $\mathbb{R}_\ell$ is defined as an $\mathbb{R}_\ell$-submodule of $\mathbb{R}_\ell^N$.

The RT weight and the RT distance over $\mathbb{R}_\ell$ are defined as follows:

**Definition 2.1** (see [5, 8]) Let $c = (c_1, c_2, \cdots, c_b) \in \mathbb{R}_\ell^b$, and

$$w_{RT}(c) = \begin{cases} max\{i : c_i \neq 0\} & c \neq 0, \\ 0 & c = 0. \end{cases}$$

The RT metric is defined by $d_{RT}(c_1, c_2) = w_{RT}(c_1, c_2)$, where $c_1, c_2 \in \mathbb{R}_\ell^b$. $w_{RT}(c)$ is called the RT weight of $c$.

**Definition 2.2** (see [17]) A spotty byte error is defined as $t$ or fewer bits errors within a $b$-bit byte, where $1 \leq t \leq b$. When none of the bits in a byte is in error, we say that no spotty byte error has occurred.

An $s$-spotty byte error is defined as a random $t$-bit error within a byte. If there are more than $t$-bit errors in a byte, the errors are defined as $m$-spotty byte errors. We can define the $m$-spotty RT weight and the $m$-spotty RT distance over $\mathbb{R}_\ell$ as follows.

**Definition 2.3** Let $e \in \mathbb{R}_\ell^N$ be an error vector and $e_i \in \mathbb{R}_\ell^b$ be the $i$-th byte of $e$, where $N = nb$ and $1 \leq i \leq n$. The number of $t/b$-errors in $e$, denoted by $w_{MRT}(e)$, and called $m$-spotty RT weight is defined as

$$w_{MRT}(e) = \sum_{i=0}^n \left\lceil \frac{w_{RT}(e_i)}{t} \right\rceil,$$

where $\lceil x \rceil$ denotes the smallest integer not less than $x$. If $t = 1$, this weight, defined by $w_{MRT}$, is equal to the $RT$ weight.

**Definition 2.4** Let $c$ and $v$ be codewords of $m$-spotty byte error control code $C$ over $\mathbb{R}_\ell$. Here $c_i$ and $v_i$ are the $i$-th bytes of $c$ and $v$, respectively. Then, $m$-spotty RT distance function between $c$ and $v$, denoted by $d_{MRT}$, is defined as follows:

$$d_{MRT}(c, v) = \sum_{i=0}^n \left\lceil \frac{d_{RT}(c_i, v_i)}{t} \right\rceil.$$

**Remark 2.5** Similar to the proof of Theorem 2.5 in [7], $m$-spotty RT distance over $\mathbb{R}_\ell$ is a metric, that is, this function satisfies the metric axioms.

## 3  The MacWilliams identity over finite commutative Frobenius rings

Hereinafter, codes will be taken to be of length $N$ where $N$ is a multiple of byte length $b$, i.e. $N = nb$.

Let $c = (c_1, c_2, \cdots, c_N)$ and $v = (v_1, v_2, \cdots, v_N)$ be two elements of $\mathbb{R}_\ell^N$. The inner product of $c$ and $v$, denoted by $\langle c, v \rangle$, is defined as follows:

$$\langle c, v \rangle = \sum_{i=1}^n \langle c_i, v_i \rangle = \sum_{i=1}^n \Big( \sum_{j=1}^b c_{(i,j)} v_{(i,b-j+1)} \Big).$$

Here, $\langle c_i, v_i \rangle = \sum_{j=1}^b c_{(i,j)} v_{(i,b-j+1)}$ denotes the inner product of $c_i$ and $v_i$, respectively. Also $c_{(i,j)}$ and $v_{(i,b-j+1)}$ are the $j$-th bits of $c_i$ and $v_i$, respectively. The inner product for each byte is taken in reverse order similar to the RT case where $n = 1$.

Now we recall some examples of finite commutative Frobenius rings and their generating characters, most of them can be found in [1] and [18].

**Remark 3.1** Here are a number of examples of finite commutative Frobenius rings.

(i) Let $R = \mathbb{F}$ be a finite field. A generating character $\chi$ on $R = \mathbb{F}$ is given by $\chi(x) = \xi^{Tr(x)}$, where $\xi = e^{\frac{2\pi i}{p}}$ and $\text{Tr} : \mathbb{F}_\ell \to \mathbb{F}_p$ is the trace function from $\mathbb{F}$ to $\mathbb{F}_p$.

(ii) Let $R = \mathbb{Z}_\ell$. Set $\xi = e^{\frac{2\pi i}{\ell}}$. Then $\chi(x) = \xi^x$, $x \in \mathbb{Z}_\ell$, is a generating character.

(iii) The finite direct sum of Frobenius rings is Frobenius. If $R_1, \cdots, R_n$ each has generating characters $\chi_1, \cdots, \chi_n$, then $R = \oplus R_i$ has generating character $\chi = \prod \chi_i$.

(iv) Any Galois ring is Frobenius. A Galois ring $R = GR(p^n, r) \cong \mathbb{Z}_{p^n}[x]/\langle f \rangle$ is a Galois extension of $\mathbb{Z}_{p^n}$ of degree $r$, where $f$ is a monic irreducible polynomial in $\mathbb{Z}_{p^n}[x]$ of degree $r$. Because $f$ is monic, any element $a$ of $R$ is represented by a unique polynomial $r = \sum_{i=1}^{r-1} a_i x^i$, with $a_i \in \mathbb{Z}_{p^n}$. Set $\xi = e^{\frac{2\pi i}{p^n}}$. Then $\chi(a) = \xi^{a_{r-1}}$.

(v) Any finite chain ring is Frobenius. Let $R$ be a finite chain ring with maximal ideal $\langle u \rangle$ and let its residue field $R/\langle u \rangle$ be $\mathbb{F}_{p^n}$, i.e, $R = \mathbb{F}_q + u\mathbb{F}_q + \cdots + u^{k-1}\mathbb{F}_q$. Any element $r$ of $R$ is represented by a unique polynomial $r = \sum_{i=1}^{r-1} a_i u^i$, with $r_i \in \mathbb{F}_q$. Set $\xi = e^{\frac{2\pi i}{q}}$. Then $\chi_r = \xi^{a_{r-1}}$.

(vi) $R = \mathbb{F}_2[u_1, u_2, \cdots, u_k]/\langle u_i^2 = 0, u_i u_j = u_j u_i \rangle$ is a Frobenius ring. Let $r_k = \sum_{A \subseteq \{1,2,\cdots,k\}} c_A u_A \in R$. Then $(c_A)$ can be thought of as a binary vector of length $2^k$. Let $wt(c_A)$ be the Hamming weight of this vector. Then $\chi(r_k) = (-1)^{wt(c_A)}$.

In order to prove our main theorem, we should first prove the following two lemmas. From now onwards, we assume $\chi$ be a generating character over finite commutative Frobenius rings in Remark 3.1. $\ell$ denotes the cardinality of $\mathbb{R}_\ell$, i.e, $|\mathbb{R}_\ell| = \ell$.

**Lemma 3.1** Let $c = (c_1, \ldots, c_b) \in \mathbb{R}_\ell^b$ with $w_{RT}(c) = j$. For any $0 \leq k \leq b$, we have

$$S^{(\ell)}(k, j) := \sum_{w_{RT}(v) = k} \chi_c(v) = \begin{cases} 1, & \text{if } k = 0; \\ \ell^{k-1}(\ell - 1), & \text{if } 1 \leq k \leq b - j; \\ -\ell^{k-1}, & \text{if } k = b + 1 - j; \\ 0, & \text{if } k \geq b + 2 - j, \end{cases}$$

where $P_k = \{v \in \mathbb{R}_\ell^b : w_{RT}(v) = k\}$.

*Proof.* It is easy to verify the result when $k = 0$. Let us assume $1 \leq k \leq b$ from here.

$$\begin{aligned} S^{(\ell)}(k,j) &= \sum_{w_{RT}(v)=k} \chi(\langle c, v \rangle) \\ &= \sum_{w_{RT}(v)=k} \chi(v_1 c_b + \cdots + v_k c_{b+1-k}) \\ &= \left( \prod_{i=1}^{k-1} \left( \sum_{v_i \in \mathbb{R}_\ell} \chi(c_{b+1-i} v_i) \right) \right) \times \left( \sum_{v_k \in \mathbb{R}_\ell^*} \chi(c_{b+1-k} v_k) \right). \end{aligned}$$

Denote $T_i = \sum_{v_i \in \mathbb{R}_\ell} \chi(c_{b+1-i} v_i)$ ($1 \leq i \leq k-1$), and $T_k = \sum_{v_k \in \mathbb{R}_\ell^*} \chi(c_{b+1-k} v_k)$. If $k \leq b - j$, then we have $c_b = \cdots = c_{b+1-k} = 0$ and

$$T_i = \sum_{v_i \in \mathbb{R}_\ell} \chi(0 \cdot v_i) = \sum_{v_i \in \mathbb{R}_\ell} 1 = \ell, \ T_k = \sum_{v_k \in \mathbb{R}_\ell^*} 1 = \ell - 1.$$

Hence $S^{(\ell)}(k,j) = \ell^{k-1}(\ell - 1)$. If $k = b + 1 - j$, we get $c_b = \cdots = c_{b+2-k} = 0$ and $c_{b+1-k} = c_j \neq 0$, and then

$$T_i = \ell, \ T_k = \sum_{v_k \in \mathbb{R}_\ell} \chi(c_j v_k) - \chi(c_j \cdot 0) = -1.$$

Hence, $S^{(\ell)}(k,j) = -\ell^{k-1}$. If $k \geq b + 2 - j$, then the last $k - 1$ positions of codeword $c$ contain at least one nonzero element, suppose for some $j \geq b - k + 2$, $c_j \neq 0$, we have

$$T_{b+1-j} = \sum_{v_{b+1-j} \in \mathbb{R}_\ell} \chi(c_j v_{b+1-j}) = 0.$$

Hence $S^{(\ell)}(k,j) = 0$. This proves the lemma.

**Lemma 3.2** Let $c = (c_1, c_2, \cdots, c_b)$ and $v = (v_1, v_2, \cdots, v_b)$ be two elements of $\mathbb{R}_\ell^b$, with $w_{RT}(c) = j$. Then $\sum_{v \in \mathbb{R}_\ell^b} \chi_c(v) z^{\lceil w_{RT}(v)/t \rceil} = V_j^{(t,\ell)}(z)$, where $V_j^{(t,\ell)}(z) = \sum_{k=0}^{b} S^{(\ell)}(k,j) z^{\lceil k/t \rceil}$.

*Proof.* Using Lemma 3.1, we can obtain

$$\begin{aligned} \sum_{v \in \mathbb{R}_\ell^b} \chi_c(v) z^{\lceil w_{RT}(v)/t \rceil} &= \sum_{k=0}^{b} \sum_{w_{RT}(v)=k} \chi_c(v) z^{\lceil k/t \rceil} \\ &= \sum_{k=0}^{b} z^{\lceil k/t \rceil} \left( \sum_{w_{RT}(v)=k} \chi_c(v) \right) \\ &= \sum_{k=0}^{b} S^{(\ell)}(k,j) z^{\lceil k/t \rceil} = V_j^{(t,\ell)}(z). \end{aligned}$$

This proves the Lemma.

Let $(G, +)$ be a finite abelian group and $V$ be a vector space over the complex numbers. The set $\widehat{G}$ of all characters of $G$ forms an abelian group under pointwise miltiplication. For any function $f : G \longrightarrow V$, define its *Fourier transform* $\widehat{f} : \widehat{G} \longrightarrow V$ by

$$\widehat{f}(\pi) = \sum_{x \in G} \pi(x) f(x), \pi \in \widehat{G}.$$

Given a subgroup $H \subseteq G$, define an *annihilator* $(\widehat{G} : H) = \{\pi \in \widehat{G} : \pi(H) = 1\}$. Moreover, we have $|(\widehat{G} : H)| = |G|/|H|$.

The Poisson summation formula relates the sums of a function over a subgroup to the sum of its Fourier transform over the annihilator of the subgroup. The following Lemma can be found in [18], which plays an impormant role in deriving the MacWilliams identity for $m$-spotty RT weight.

**Lemma 3.3** (Poisson Summation Formula) Let $H \subset G$ be a subgroup, and let $f : G \longrightarrow V$ be any function from $G$ to a complex vector space $V$. Then

$$\sum_{x \in H} f(x) = \frac{1}{|(\widehat{G} : H)|} \sum_{\pi \in (\widehat{G}:H)} \widehat{f}(\pi).$$

Let $\alpha_j = \#\{i : w_{RT}(c_i) = j, 1 \leq i \leq n\}$. That is, $\alpha_j$ is the number of bytes having RT weight $j$, $0 \leq j \leq b$, in a codeword. The summation of $\alpha_0, \alpha_1, \cdots, \alpha_b$ is equal to the code length in bytes, that is $\sum_{j=0}^{b} \alpha_j = n$. The RT weight distribution vector $(\alpha_0, \alpha_1, \cdots, \alpha_b)$ is determined uniquely for the codeword $c$. Then, the $m$-spotty RT weight of the codeword $c$ is expressed as $w_{MRT}(c) = \sum_{j=0}^{b} \lceil j/t \rceil \cdot \alpha_j$. Let $A_{(\alpha_0, \alpha_1, \cdots, \alpha_b)}$ be the number of codewords with RT weight distribution vector $(\alpha_0, \alpha_1, \cdots, \alpha_b)$. For example, let (010 012 020 202 000 200) be a codeword over $\mathbb{F}_3$ with byte 3. Then, the RT weight distribution vector of the codeword is $(\alpha_0, \alpha_1, \alpha_2, \alpha_3) = (1, 1, 2, 2)$. Therefore, $A_{(1,1,2,2)}$ is the number of codewords with RT weight distribution vector $(1, 1, 2, 2)$.

We are now ready to define the $m$-spotty RT weight enumerator of a byte error control code over $\mathbb{R}_\ell$.

**Definition 3.2** The weight enumerator for $m$-spotty byte error control code $C$ is defined as

$$W(z) = \sum_{c \in C} z^{w_{MRT}(c)}.$$

By using the parameter $A_{(\alpha_0, \alpha_1, \cdots, \alpha_b)}$, which denotes the number of codewords with RT weight distribution vector $(\alpha_0, \alpha_1, \cdots, \alpha_b)$, $W(z)$ can be expressed as follows:

$$W(z) = \sum_{\substack{(\alpha_0, \ldots, \alpha_b) \\ \alpha_0, \ldots, \alpha_b \geq 0 \\ \alpha_0 + \cdots + \alpha_b = n}} A_{(\alpha_0, \ldots, \alpha_b)} \prod_{j=0}^{b} (z^{\lceil j/t \rceil})^{\alpha_j}.$$

The next theorem holds for the weight enumerator $W(z)$ of the code and that of the dual code $C^\perp$, expressed as $W^\perp(z)$.

**Theorem 3.1** Let $C$ be a linear code and $C^\perp$ be its dual code. The relation between the $m$-spotty RT weight enumerators of $C$ and $C^\perp$ is given by

$$W^\perp(z) = \sum_{\substack{(\alpha_0, \ldots, \alpha_b) \\ \alpha_0, \ldots, \alpha_b \geq 0 \\ \alpha_0 + \cdots + \alpha_b = n}} A^\perp_{(\alpha_0, \ldots, \alpha_b)} \prod_{j=0}^{b} (z^{\lceil j/t \rceil})^{\alpha_j} = \frac{1}{|C|} \sum_{\substack{(\alpha_0, \ldots, \alpha_b) \\ \alpha_0, \ldots, \alpha_b \geq 0 \\ \alpha_0 + \cdots + \alpha_b = n}} A_{(\alpha_0, \ldots, \alpha_b)} \prod_{j=0}^{b} (V_j^{(t,\ell)}(z))^{\alpha_j}.$$

*Proof.* Given a linear code $C \subset \mathbb{R}_\ell^n$, we apply the Poisson Summation Formula with $G = \mathbb{R}_\ell^n$, $H = C$, and $V = \mathbb{C}[z]$, the polynomial ring over $\mathbb{C}$ in one indeterminate. The first task is to identify the character-theoretic annihilator $(\widehat{G} : H) = (\widehat{\mathbb{R}_\ell}^n : C)$ with $C^\perp$. Let $\rho$ be a generating character of $\mathbb{R}_\ell$. We use $\rho$ to define a homomorphism $\beta : \mathbb{R}_\ell \longrightarrow \widehat{\mathbb{R}_\ell}$. For $r \in \mathbb{R}_\ell$, the character $\beta(r) \in \widehat{\mathbb{R}_\ell}$ has the form $\beta(r)(s) = (r\rho)(s) = \rho(sr)$ for $s \in \mathbb{R}_\ell$. One can verify that $\beta$ is an isomorphism of $\mathbb{R}_\ell$-modules. In particular, wt$(r) =$ wt$(\beta r)$, where wt$(r) = 0$ for $r = 0$, and wt$(r) \neq 0$ for $r \neq 0$.

Extend $\beta$ to an isomorphism $\beta : \mathbb{R}_\ell^n \longrightarrow \widehat{\mathbb{R}_\ell}^n$ of $R$-modules, via $\beta(x)(y) = \rho(yx)$, for $x, y \in \mathbb{R}_\ell^n$. Again, $w_{RT}(x) = w_{RT}(\beta x)$. For $x \in \mathbb{R}_\ell^n$, $\beta(x) \in (\widehat{\mathbb{R}_\ell}^n : C)$ means $\beta(x)(C) = \beta(C \cdot x) = 1$. This means that the ideal $C \cdot x$ of $\mathbb{R}_\ell$ is contained in ker$(\rho)$. Because $\rho$ is a generating character, which implies that $C \cdot x = 0$. Thus $x \in C^\perp$. The converse is obvious. Thus $C^\perp$ corresponds to $(\widehat{\mathbb{R}_\ell}^n : C)$ under the isomorphism $\beta$.

Remember that $\beta : \mathbb{R}_\ell^n \longrightarrow \widehat{\mathbb{R}_\ell}^n$ is an isomorphism of $R$-modules and $(C^\perp)^\perp = C$. Thus the Poisson Summation Formula becomes

$$\sum_{v \in C^\perp} f(v) = \frac{1}{|C|} \sum_{c \in C} \widehat{f}(c),$$

where the Fourier transform is
$$\widehat{f}(c) = \sum_{v \in \mathbb{R}_\ell^N} \chi_c(v) f(v).$$

Define $f(v) = \prod_{i=1}^n z^{\lceil w_{RT}(v_i)/t \rceil}$, where $v_i$ denotes the $i$-th bytes of $v$. Then we can get

$$\begin{aligned}
\widehat{f}(c) &= \sum_{v \in \mathbb{R}_\ell^N} \chi_c(v) \prod_{i=1}^n z^{\lceil w_{RT}(v_i)/t \rceil} \\
&= \sum_{v \in \mathbb{R}_\ell^{nb}} \prod_{i=1}^n \chi_{c_i}(v_i) \prod_{i=1}^n z^{\lceil w_{RT}(v_i)/t \rceil} \\
&= \prod_{i=1}^n \left( \sum_{v_i \in \mathbb{R}_\ell^b} \chi_{c_i}(v_i) z^{\lceil w_{RT}(v_i)/t \rceil} \right) \\
&= \prod_{i=1}^n V_{w_{RT}(c_i)}^{(t,\ell)}(z).
\end{aligned}$$

Assume that RT weight of the fixed vector $c_i$ is $w_{RT}(c_i) = j$, and $c$ has the RT weight distribution vector $(\alpha_0, \ldots, \alpha_b)$, then we have

$$\widehat{f}(c) = \prod_{j=0}^b (V_j^{(t,\ell)}(z))^{\alpha_j}.$$

Thus we have

$$\sum_{c \in C^\perp} \prod_{j=0}^b (z^{\lceil j/t \rceil})^{\alpha_j} = \frac{1}{|C|} \sum_{c \in C} \prod_{j=0}^b (V_j^{(t,\ell)}(z))^{\alpha_j}.$$

After rearranging the summations on both sides according to the RT weight distribution vectors of codewords in $C^\perp$ and $C$ respectively, we have the result

$$\sum_{\substack{(\alpha_0,\ldots,\alpha_b) \\ \alpha_0,\ldots,\alpha_b \geq 0 \\ \alpha_0+\cdots+\alpha_b=n}} A_{(\alpha_0,\ldots,\alpha_b)}^\perp \prod_{j=0}^b (z^{\lceil j/t \rceil})^{\alpha_j} = \frac{1}{|C|} \sum_{\substack{(\alpha_0,\ldots,\alpha_b) \\ \alpha_0,\ldots,\alpha_b \geq 0 \\ \alpha_0+\cdots+\alpha_b=n}} A_{(\alpha_0,\ldots,\alpha_b)} \prod_{j=0}^b (V_j^{(t,\ell)}(z))^{\alpha_j}.$$

This proves the main results.

## 4 Examples

In Section 3, we present a proof of MacWilliams identities that is valid over any finite commutative Frobenius ring. In this section, we take two examples to illustrate our main results.

**Example 4.1** Let
$$G = \begin{pmatrix} 1 & 0 & 2 & 2 & 2 & 0 & 1 & 0 & 0 \\ 0 & 1 & 1 & 0 & 1 & 0 & 0 & 0 & 0 \end{pmatrix}$$
be the generator matrix of a ternary linear code $C$ of length 9. $C$ has 9 codewords. The dual code of $C$ is a ternary linear code of length 9 and it has 2187 codewords.

Before computing the $m$-spotty weight enumerator of $C$, we illustrate how to apply the formulae. It is easy to show that the codeword $c = (011\ 010\ 000)$ belongs to $C$. Let $b = 3$ and $t = 2$. Then, the RT weight distribution vector of the codeword is $(\alpha_0, \alpha_1, \alpha_2, \alpha_3) = (1, 0, 1, 1)$. The RT weight distribution vectors of the codewords of $C$, the number of codewords, and polynomials $V_j^{(t,\ell)}$ for $b = 3$ and $t = 2$ are shown in Tables I and II for the necessary computations to apply Theorem 3.1.

**Table I**
RT weight distribution vectors of the codewords
in $C$ and the number of codewords.

| RT weight vector | number |
|---|---|
| $(3,0,0,0)$ | 1 |
| $(0,1,1,1)$ | 4 |
| $(1,0,1,1)$ | 2 |
| $(0,2,1,0)$ | 2 |

**Table II**
Polynomials $V_j^{(2,3)}$ for $t=2$ and $b=3$.

| |
|---|
| $V_0^{(2,3)}(z) = 1 + 8z + 18z^2$ |
| $V_1^{(2,3)}(z) = 1 + 8z - 9z^2$ |
| $V_2^{(2,3)}(z) = 1 - z$ |
| $V_3^{(2,3)}(z) = 1 - z$ |

According to the expression of $W(z)$ and Table I, we obtain the $m$-spotty weight enumerator of $C$ as $W(z) = 1 + 4z^3 + 4z^4$. By applying Theorem 3.1 and Table II, we obtain

$$\begin{aligned} W^\perp(z) &= \frac{1}{|C|} \sum_{\alpha_0+\alpha_1+\alpha_2+\alpha_3=3} A_{(\alpha_0,\alpha_1,\alpha_2,\alpha_3)} \prod_{j=0}^{b} (V_j^{(2,3)}(z))^{\alpha_j} \\ &= \frac{1}{9}(V_0^{(2,3)}(z))^3 + \frac{4}{9}(V_1^{(2,3)}(z))(V_2^{(2,3)}(z))(V_3^{(2,3)}(z)) \\ &\quad + \frac{2}{9}(V_0^{(2,3)}(z))(V_2^{(2,3)}(z))(V_3^{(2,3)}(z)) + \frac{2}{9}(V_1^{(2,3)}(z))^2(V_3^{(2,3)}(z)) \\ &= 1 + 10z + 24z^2 + 116z^3 + 542z^4 + 846z^5 + 648z^6. \end{aligned}$$

**Example 4.2** Let

$$G = \begin{pmatrix} 1 & 1 & 1 & 5 & 4 & 2 \\ 0 & 3 & 0 & 3 & 3 & 3 \\ 0 & 0 & 3 & 3 & 0 & 3 \end{pmatrix}$$

be the generator matrix of a linear code $C$ over $\mathbb{Z}_6$ of length 3. $C$ has 24 codewords. The dual code of $C$ is also a linear code of length 6 and it has 1944 codewords.

The number of codewords, and polynomials $V_j^{(t,\ell)}$ for $b=3$ and $t=2$ are shown in Tables III and IV for the necessary computations to apply Theorem 3.1.

**Table III**
RT weight distribution vectors of the codewords
in $C$ and the number of codewords.

| RT weight vector | number |
|---|---|
| $(2,0,0,0)$ | 1 |
| $(0,0,0,2)$ | 18 |
| $(0,1,0,1)$ | 1 |
| $(0,0,1,1)$ | 3 |
| $(0,1,1,0)$ | 1 |

According to the expression of $W(z)$ and Table III, we obtain the $m$-spotty weight enumerator of $C$ as

$$W(z) = 1 + z^2 + 4z^3 + 18z^4.$$

**Table IV**
Polynomials $V_j^{(2,6)}$ for $t = 2$ and $b = 3$.

| |
|---|
| $V_0^{(2,6)}(z) = 1 + 35z + 180z^2$ |
| $V_1^{(2,6)}(z) = 1 + 35z - 36z^2$ |
| $V_2^{(2,6)}(z) = 1 - z$ |
| $V_3^{(2,6)}(z) = 1 - z$ |

By applying Theorem 3.1 and Table IV, we obtain

$$\begin{aligned}
W^{\perp}(z) &= \frac{1}{|C|} \sum_{\alpha_0+\alpha_1+\alpha_2+\alpha_3=3} A_{(\alpha_0,\alpha_1,\alpha_2,\alpha_3)} \prod_{j=0}^{b} (V_j^{(2,6)}(z))^{\alpha_j} \\
&= \frac{1}{24} [(V_0^{(2,6)}(z))^2 + 18(V_3^{(2,6)}(z))^2 + (V_1^{(2,6)}(z))(V_2^{(2,6)}(z)) + (V_1^{(2,6)}(z))^2(V_3^{(2,6)}(z)) \\
&\quad + 3(V_2^{(2,6)}(z))^2(V_3^{(2,6)}(z))] \\
&= \frac{1}{24} [(1 + 35z + 180z^2)^2 + 18(1-z)^2 + 2(1 + 35z - 36z^2)(1-z) + 3(1-z)^2] \\
&= 1 + 4z + 61z^2 + 528z^3 + 1350z^4.
\end{aligned}$$

## 5 Conclusion

In this paper, we prove the MacWilliams identities for $m$-spotty RT weight enumerators over arbitrary finite commutative Frobenius rings. We conclude the paper by giving two illustrations of the theorems. This provides the relation between the $m$-spotty RT weight enumerator of the code and that of the dual code. Also, the indicated identities include the MacWilliams identity over $\mathbb{F}_2$ in [7] as a special case.

## 6 Acknowledgments

This research was done while the author was visiting CCRG of Nanyang Technological University. The author wishes to thank Profesor San Ling for helpful discussion.